# Chromatic Effect for THz Generation in a Novel Wave-front Tilt Scheme

Bin Li[1,3]*, Wenyan Zhang[1], Xiaoqing Liu[1], Jianhui Chen[1], Haixiao Deng[1], Chao Feng[1], Lie Feng[1], Taihe Lan[1], Bo Liu[1], Jia Liu[5], Dong Wang[1,3], Xingtao Wang[1], Zhinan Zeng[2]*, Iijian Zhang[4], Tong Zhang[1], Zhentang Zhao[1,3]

[1]*Shanghai Institute of Applied Physics, Chinese Academy of Sciences, 2019 Jialuo Road, Shanghai 201800, China*

[2]*Laboratory for High Intensity Optics, Shanghai Institute of Optics and Fine Mechanics, Chinese Academy of Sciences, 390 Qinghe Road, Shanghai 201800, China*

[3]*School of Physical Science and Technology, ShanghaiTech University, 100 Haike Road, Shanghai 200031, China*

[4]*National Laboratory of Solid State Microstructures and College of Engineering and Applied Sciences, Nanjing University, Nanjing 210093, China*

[5]*European XFEL GmbH, Holzkoppel 4, 22869 Schenefeld, Germany*

*Corresponding Authors:
libin1995@sinap.ac.cn & zhinan_zeng@mail.siom.ac.cn

## Abstract

Deriving single or few cycle terahertz (THz) pulse by intense femtosecond laser through cascaded optical rectification in electro-optic crystals is a crucial technique in cutting-edge time-resolved spectroscopy to characterize micro-scale structures and ultrafast dynamics. In the past decade, lithium niobate (LN) crystal implementation of wave-front tilt scheme has been prevalently used, while painstaking efforts have been invested in order to achieve higher THz conversion efficiency. In this research we developed a brand new type of LN crystal possessing dual-face-cut and Brewster's coupling, and conducted experimental and simulative investigation systematically to optimize the multi-dimensionally entangled parameters in THz generation, predicting the extreme conversion efficiency of ~10% is potentially promising at the THz absorption coefficient of ~0.5cm$^{-1}$. More remarkably, we first discovered that the chirp of the driving laser pulse plays a decisive role in the wave-front tilt scheme, and the THz generation efficiency could be enhanced tremendously by applying an appropriate chirp.

## I. Introduction

THz wave spans the frequency range of 0.1-10 THz, associated with wavelength of 30micron-3mm, representing the most extraordinary characteristics in the overall electro-magnetic spectrum. During the past decade, the THz technology and applications are developing rapidly, widely used in material characterization, molecular imaging and structural tomography, investigation of electron-phonon interaction and dynamics, non-destructive bio-medical diagnostics etc. [1, 2, 3, 4, 5, 6, 7, 8]. Numerously novel ideas and delicately experimental techniques are proposed and established to generate and detect THz waves [9, 10, 11]. Among them, the optical rectification has been demonstrated to create THz pulses most efficiently, through applying the ultrafast laser to induce different frequency generation (DFG) in the THz photonic crystal [12, 13, 14, 15, 16]. Recent years, lithium niobate (LN) crystal, either in its stoichiometric or congruent form turns into preferable THz generation media due to its large optical-to-THz conversion efficiency, compared to many others, e.g. GaAs, ZnTe, GaP and etc. [17, 18, 19]. However an instant drawback of using LN is that its optical refractive index of far-infrared spectrum (including THz) is very different with that of near-infrared and visible band. Thus the phase-matching in-between the driving laser and THz wave is not straight forward achieved, which limits the THz generation efficiency. Cherenkov radiation scheme is then adopted and demonstrated to enhance the THz wave generation efficiency remarkably, where the laser pulse front is tilted purposely with respect to its propagation direction [20, 21, 22, 23, 24]. Thus DFG takes place within the intra-band of the laser pulse through a non-collinear geometry, where the generated THz wave is expected to propagate towards the wave-front of the driving pulse.

THz pulse generation implementation of the wave-front tilt scheme through DFG process in LN is illustrated in Fig.1 (and the inserted vector diagram), where a photon of the driving laser $\vec{k}(\omega_L)$ is decomposed into a THz photon $\vec{k}(\Omega_{THz})$ and an idler photon $\vec{k}(\omega_L - \Omega_{THz})$. And the process could be cascaded multiple times (refer to Eq. (1)), so the final photon energy of the residual driving beam would be significantly smaller than its initial value (up to a few tens of terahertz), displaying apparent "red-shift" broadening spectral features (refer to Fig.5) [21, 25].

$$\begin{cases} \vec{k}(\omega_L) = \vec{k}(\omega_L - \Omega_{THz}) + \vec{k}(\Omega_{THz}) \\ \vec{k}(\omega_L - \Omega_{THz}) = \vec{k}(\omega_L - 2\Omega_{THz}) + \vec{k}(\Omega_{THz}) \\ \vec{k}(\omega_L - 2\Omega_{THz}) = \vec{k}(\omega_L - 3\Omega_{THz}) + \vec{k}(\Omega_{THz}) \\ \quad \ldots \ldots \end{cases} \quad (1).$$

The phase (or wave vector) match condition in-between the driving laser and THz wave through the non-collinear coupling scheme in Fig.1 could be used to determine the optimal wave-front tilt angle for the driving beam in LN,

$$\Delta \vec{k} = \vec{k}(\Omega_{THz}) + \vec{k}(\omega_L - \Omega_{THz}) - \vec{k}(\omega_L) \quad (2).$$

Thus,

$$\left| \Delta \vec{k} \right| \approx \frac{\Omega_{THz} \cdot n(\Omega_{THz})}{c} \cos\theta_{LN} - \left(\frac{dk}{d\omega}\right)_L \cdot \Omega_{THz} = \frac{\Omega_{THz}}{c}\left[ n(\Omega_{THz})\cos\theta_{LN} - n_g(\omega_L) \right] = 0 \quad (3),$$

which leads to

$$\theta_{LN} = \cos^{-1}\left( \frac{n_g(\omega_L)}{n(\Omega_{THz})} \right) \quad (4).$$

Where $n_g(\omega_L)$ and $n(\Omega_{THz})$ are the group index of the driving laser, and the phase index of THz wave in LN respectively. As shown in Fig.1, the optimal wave-front tilt angle $\theta_{LN}$ is actually the desired intersect angle in-between the propagation directions of the laser and THz wave. While the THz wave propagates a distance of L, parallel to the wave-front normal of the driving laser, the laser pulse would propagate a distance of $L/\cos\theta_{LN}$ in the meantime [23, 26].

The rest of this article is organized as following. Section II introduces our self-made THz generation/characterization setup utilizing a unique wave-front tilt scheme. Section III describes the simulations and strategies for optimal THz generation. Section IV outlines the experimental results along with the relevant analysis and discussions. In section V, we finish with general remarks and conclusions for our study.

## II. Setup and Apparatus

The THz experimental setup is already presented in Fig. 1 schematically. Compared to the typical wave-front tilt scheme previously used for THz generation in LN, through applying pump laser beam with vertical polarization (s-pol) to incident on a side plane of a horizontal 'z-cut' crystal [19, 20, 21, 22, 23, 24, 27], the major difference and advantage of our setup is that the LN crystal is in presence of both **horizontal** 'x-cut' and **vertical** 'z-cut', where a laser beam with horizontal polarization (p-pol) incidents onto a side plane, approaching Brewster's angle to reduce the Fresnel loss at the 'air-LN' interface. Since the material chromatic dispersion in LN is relatively small around ~1μm, so the design is potentially applicable for a broad driving spectral range with only minor tuning needed. The unique double-face-cut LN crystal is congruently doped with MgO of 6.0% molecular concentration, fabricated by (United Crystal Inc., USA) through customer request. It is shaped in a triangular prism, with dimensions of 15.5mm x 22.4mm (Y) x 32.1mm in cross-section and a height of 15.0mm (X). Upon Brewster's incidence at a side surface of 15.5mm x 15.0mm(X), the refractive laser beam would propagate along y-axis after entering into the crystal with polarization remaining parallel to the z-axis (optical axis), to ensure the THz generation efficiency.

The driving pulse is provided by a conventional Ti:Sapphire laser system (Coherent Legend Duo regenerative laser amplifier), with 1kHz repetition rate, pulse energy up to ~3.6mJ, FWHM bandwidth of ~7.6nm at 800nm (central wavelength), and pulse duration of ~130fs, corresponding to a transform-limit (TL) pulse of nearly full coherence. A down-collimation telescope (not shown) is used to reduce the radius of the laser beam to ~5mm to enhance the driving beam fluence (as well as the beam intensity). A blazing type plane grating with groove density of 2200ln/mm optimally blazed at 800nm is implemented to create the desired wave-front tilt for the driving laser beam. The p-pol laser beam incidents at 65° respected to the grating normal, thus the $1^{st}$ order diffraction beam '1' is outgoing at ~58.6°. The beam incidence-diffraction geometry is approaching 'Littrow' condition to achieve diffraction efficiency beyond ~90% (for the $1^{st}$ order), with a few percent of zero-order '0', minor other orders and loss. This geometry would also allow the diffraction beam well separate from the incidence beam, while possessing tremendous wave-front tilt in the air (73.5°). The beam is further delivered through a pair of cylindrical lenses, $CL_1$ ($f_1$ = 300mm) and $CL_2$ ($f_2$ = 100mm), which are curved horizontally but flat vertically. Thus the diffractive beam is refocused twice in its dispersive (horizontal) coordinate with moderate divergence in non-dispersive (vertical) coordinate. The well coupled lens-pair "$f_1$-$f_1$-$f_2$-$f_2$" is utilized and precisely aligned to reduce the optical aberration of the driving beam, where the front focal spot of $CL_1$ is located at the surface of the grating (G), the rear focal spot of $CL_1$ is overlapped with the front focus of $CL_2$, and the chromatic-dispersed beam would refocus at the rear focus of $CL_2$ (located inside the LN crystal). Concerning the overall transport efficiency for the driving beam of 55%, the ultimate pulse energy of ~2mJ could be coupled into the LN crystal for THz generation. The LN crystal is installed on a crystal mount with translational mechanism in "y-z plane" and rotational mechanism around "x-axis" (refer to Fig.1). The in-plane motion could change the laser propagation length in the crystal for the THz generation, and alter the driving laser beam fluence through adjusting the position of LN back and forth towards the focal spot of the chromatic beam. The rotation could fine adjust the crystal orientation in y-z plane to achieve the desired phase match angle. Therefore, the coordinates of LN could be well optimized for the THz generation at various driving pulse parameters.

Upon incidence into the LN crystal, the refractive beam would possess a wave-front tilt angle of ~64°, satisfying the optimal phase match in-between the driving laser (~800nm) and expected THz beam (~1THz). The THz pulse would propagate along the wave-front of the driving beam, and eventually exit from another side plane 32.1mm x 15.0mm (X) of the LN crystal in normal, where its pulse energy is measured by a pyroelectric detector with proper THz filters, and cross-calibrated by a Golay Cell (both from Microtech Instrument Inc., USA). An optical beam chopper operated in the chopping rate of 10-50Hz provides periodically electric modulation for the THz pulse energy calibration device. The temporal waveform of THz pulse is detected by a home-made electro-optical (EO) sampling apparatus separately [11, 14]. The THz beam is refocused by a couple of 90° off-axis parabola (OAP) to achieve the best focal spot in a ZnTe crystal with thickness of 0.5-2mm, where the 2$^{nd}$ OAP is centrally holed to allow the probe beam passing through the parabola and overlapping with the THz pulse at the ZnTe crystal. The optical probe beam originates from the '0'-order component of the grating, passing through a delay-scan stage to change the mutual delay in-between the probe and THz pulses. After penetrating the ZnTe crystal, the probe beam sequentially transmits a best-form focal lens, a quarter wave-plate (λ/4), a Wollaston prism, then splits and finally hits into a balanced photo-diode (PD). The EO signal from the balanced PD is phase-retrieved and recorded by a lock-in amplifier which is triggered by the synchronization signal of the laser system at 1 kHz repetition rate. Therefore the EO signal is sampled at the single-shot mode without applying beam chopper. In a typical delay scan EO-sampling measurement, we set the scan-delay-step as 1.5μm and 1024 steps in total, associated with the temporal resolution of ~10fs, and a timing window of ~10ps, which could characterize the THz waveform with high temporal resolution. The typical integration time at each delay scan step is few tenth of a second, which means each data point is achieved by averaging a few hundred of single-shots to provide decent S/N ratio. It takes ~10 minutes to complete a full delay-scan THz field measurement.

### III. Simulation and Optimization
**A) Lens-pair coupling for various grating groove densities**

The desired wave-front tilt of the driving beam is created by a diffractive grating with groove line spacing of $d_0$ through the 1$^{st}$ order diffraction. As illustrated in Fig. 2(a), a blazing type grating optimally blazed at the central wavelength $\lambda_c \sim 800nm$ (i.e. $\theta_{BL} = \theta_{Litt} \simeq \sin^{-1}\left[\dfrac{\lambda_c}{2d_0}\right]$) is used, working at the proposed incidence angle of $\beta_0 \approx \theta_{Litt} + 3°$, thus the diffraction beam (diffracted at the angle of $\beta_1 \approx \sin^{-1}\left[\frac{\lambda_c}{d_0} - \sin\beta_0\right]$) would present a tilt angle for its pulse-front, according to

$$\tan\theta_G = \lambda \frac{d\beta}{d\lambda} \approx \frac{\lambda_c}{d_0 \cdot \cos[\beta_1(\lambda_c)]} \quad (5).$$

The beam is then delivered through a pair of de-magnification cylindrical lens (CL), with focal lengths of $f_1$ and $f_2$ respectively in Tangential coordinate ($f_1 > f_2$). The grating, $CL_1$, $CL_2$ and LN crystal form a well coupled "$f_1$-$f_1$-$f_2$-$f_2$" imaging system. Upon the monochromatic approximation, the two cylindrical lenses could effectively reduce the diffractive beam size to $f_2/f_1$ at the rear focal spot of $CL_2$ (located inside the LN crystal). Implementing the parameters of the optical layout in Fig.1, the effective wave-front tilt angle of the driving beam in the LN crystal could eventually be described,

$$\tan\theta_{LN} = \frac{\tan\theta_G}{n_g(\lambda_c)} \frac{\cos\theta_i}{\cos\theta_t} \frac{f_1}{f_2} \quad (6).$$

Where $\theta_i$ (set to Brewster's angle $\theta_B$ at $\lambda_c$) and $\theta_t$ are the incidence and refraction angle of the laser beam at the air-LN interface. The result of $\theta_{LN}$ in Eq. (6) should be equal to that in Eq. (4), so the "phase-matched" de-magnification of $f_1/f_2$ could be determined for a series of gratings with groove densities of 1000-2400ln/mm. The results are plot in Fig.2 (b), where the incidence and diffraction angles are almost equally distributed around the Littrow angle $\theta_{Litt}$ for each grating, and the angular distribution of the diffraction beam in the figure is due to the intra-band chromatic dispersion of the laser pulse (light-blue area and left-axis); the calculated values of "$f_1/f_2$" decreases monotonously with the groove density (right axis). More specifically, the lower groove density (e.g. 1000ln/mm) would require higher de-magnification ($f_1/f_2>10$), leading to un-balanced and asymmetric optical design; while the higher groove density (e.g. 2000ln/mm) could achieve reasonably lower de-magnification coupling ($f_1/f_2\sim4$).

Concerning the original laser beam diameter $\sigma_0$ of 5mm at the upper stream of the grating (Fig.1), the effective beam size in LN crystal could be calculated for each case, where "effective" means the actual laser pulse energy density distribution in the LN crystal with consideration of the beam transverse stretching effect caused by the wave-front tilt,

$$\sigma_{LN} = \frac{\sigma_0}{\cos\theta_{LN}} \frac{\cos\beta_1}{\cos\beta_0} \frac{\cos\theta_t}{\cos\theta_i} \frac{f_2}{f_1} \quad (7).$$

Then the calculated laser beam fluencies are plot in Fig.2(c), for various gratings and within the laser pulse energy in the range of 0.1-2mJ. It clearly shows that for each laser pulse energy, the expected beam fluence decreases along with the increasing of the groove density of the grating, which is consistent with the feature of the lens-pair demagnification curve in Fig.2 (b). Eventually the grating groove density of 2200ln/mm is selected for the THz generation in experiment, where the demagnification of $f_1/f_2$ is ~3, and the driving beam fluencies in LN are expected to be moderate, 0.2-4mJ/cm$^2$. According to previous experiments of using LN, the optimal driving beam fluence of the femtosecond laser is claimed to be within the range of 1-10mJ/cm$^2$, which is not only far below the damage threshold of the crystal, but also smaller than the THz saturation regime, where the pump induced free-carriers would cause strong absorption of the THz waves [13, 17, 19, 28].

**B) Driving pulse intra-band chromatic distribution**

The discussion above is mainly concerning of a monochromatic wave, however ultrafast lasers (e.g. a conventional Ti:sapphire laser) are used in the experiment, spanning certain spectral range. The THz waves origin from the intra-band DFG process of the laser, and the chromatic dispersion and angular distribution therein could be evaluated. While the laser wave-packet passes through the dispersive elements of the system (consisting of e.g. the grating, the relay lens pair and LN), the different colors in the laser spectrum would eventually incident on the LN crystal with slightly different angles, causing the deviation of the actual wave-front tilts from the perfect phase match angle. The angular distribution of the driving beam would then lead to the declining of THz generation coherence length and the generation efficiency as well.

With including the intra-band angular distribution, the wave-front tilt generation would be wavelength dependence, but similar to Eq. (5)

$$\tan\theta_G(\lambda) = \frac{\lambda}{d_0 \cdot \cos[\beta_1(\lambda)]} = \frac{\lambda}{d_0\sqrt{1-\left(\frac{\lambda}{d_0}-\sin\beta_0\right)^2}} \qquad (8).$$

Extrapolating from Eq. (6), the realistic wave-front tilt angle in LN crystal should be,

$$\tan\theta_{LN}(\lambda) = \frac{\tan\theta_G(\lambda)}{n_g(\lambda)}\frac{\cos\theta_i(\lambda)}{\cos\theta_t(\lambda)}\frac{f_1}{f_2} \qquad (9).$$

Where the incidence angle $\theta_i(\lambda)$ and refraction angle $\theta_t(\lambda)$ at the "air-LN" interface could be calculated with respect to that of the central wavelength (i.e. set as the Brewster's angle $\theta_B$),

$$\theta_i(\lambda) = \theta_c + \Delta\theta_i(\lambda) \approx \theta_B + \Delta\theta_i(\lambda) \qquad (10).$$

As illustrated in Fig. 3(a), the incidence angle distribution at the air-LN interface $\Delta\theta_i(\lambda)$ originates from the diffraction grating, propagates through the lens pair CL$_1$ and CL$_2$, experiencing double refocusing, and finally reaches $\Delta\theta_i(\lambda) \approx \frac{f_1}{f_2}\Delta\beta_i(\lambda)$ at the para-axial approximation. With implementation of the grating formula,

$$\theta_i(\lambda) \approx \theta_B + \Delta\theta_i(\lambda) = \theta_B + \frac{(\lambda - \lambda_c)}{d_0 \frac{f_2}{f_1} \sqrt{1 - \left(\frac{\lambda}{d_0} - \sin\beta_0\right)^2}} \quad (11).$$

From (8), (9), (10), (11) and Snell's law, eventually we got

$$\tan\theta_{LN}(\lambda) = \frac{\lambda}{d_0 \frac{f_2}{f_1} \sqrt{1 - \left(\frac{\lambda}{d_0} - \sin\beta_0\right)^2}} \frac{n_f(\lambda)}{n_g(\lambda)} \frac{\sqrt{1 - (\sin\theta_i(\lambda))^2}}{\sqrt{n_f(\lambda)^2 - (\sin\theta_i(\lambda))^2}} \quad (12).$$

Where the optimal phase-match condition is presumably set at the central wavelength for convenience, and $n_f(\lambda)$ or $n_g(\lambda)$ represents the phase or group index of LN. The effective THz generation coherence length at various intra-band wave-lengths of the driving laser could be evaluated as:

$$L_{c\,o}(\lambda) = \frac{2\pi}{\Delta k(\lambda)} \approx \lambda_{THz} \left[ \frac{1}{n_{THz}(\lambda_{THz})\cos\theta_{LN}(\lambda) - n_g(\lambda)} \right] \quad (13).$$

This clearly shows that the major de-coherence factor for THz generation is due to the wave-front tilt deviation from the ideal phase match caused by the intra-band chromatic dispersion and angular distribution of the driving beam, which leads to the decreasing of the coherence length for THz generation and limits the efficiency. Plugging the relevant parameters and assuming at ~1THz, the actual wave-front tilt angle and corresponding THz generation coherence length within the intra-band of the driving laser could be calculated and plot in Fig. 3(b) as the curves scaled in the left and right axes respectively. As indicated in the figure as well, for TL driving pulse duration of ~130fs used in our experiment associated with a FWHM bandwidth of ~7.6nm, the effective coherent length for THz generation is roughly about ~2mm.

In Fig. 3(c), we plot a series of THz generation coherence length by applying various full coherence driving pulse durations from ~50fs up to ~1ps, where the TL intra-band distribution of each driving pulse is concerned, then the corresponding chromatic dispersion, wave-front tilt distribution and coherence generation length could be calculated according to Eq. (11-13). And the curve exhibits that the coherent length is less than 2mm for using driving pulse length of <100fs, but beyond 10mm for using pulse width of >500fs.

C) **Optimal driving pulse duration & THz generation length**
In previous section, the correlation in-between "coherence length for THz generation" and

"driving pulse duration" is investigated, it seems that a longer pulse is more favorable in THz generation, since a short pulse associated with broader TL bandwith causes larger chromatic dispersion and angular distribution, which limits the THz generation. However in fact, the optimal driving pulse is certainly not as long as possible, since the THz wave originates from the intra-band cascaded DFG process, a bandwidth spanning would be a prerequisite to initialize the DFG reaction chain. And in a typical DFG process, the conversion efficiency is typically proportional to the peak intensity of the driving laser before achieving saturation (i.e. in linear conversion region). So the optimal driving pulse duration should be determined through compromising these two opposite but competitive directions. In the meantime, the THz absorption in the generation media should also be considered, since the large optical rectification coefficient of the THz generation media is unavoidably associated with large probability for the reverse process, corresponding to strong absorption. Besides that, the driving pulse induced free-carriers in the media are expected to absorb the THz wave substantially. So in our experiment, we only apply moderate driving beam fluence, within the region of few mJ/cm$^2$ to avoid saturation and carrier induced absorption. The conversion efficiency could be numerically simulated by using Eq. (14),

$$\eta_{THz} = \frac{2\Omega_{THz}^2 d_{eff}^2 L_{THz}^2}{\varepsilon_0 n(\omega_L)^2 n(\Omega_{THz}) c^3} \frac{F}{\tau} \frac{\sin^2\left[\Delta k(\tau) \cdot L_{THz}/2\right]}{\left[\Delta k(\tau) \cdot L_{THz}/2\right]^2} \exp[-\alpha_{THz} L_{THz}] G(L_{THz}, \tau) \quad (14).$$

Where the phase matching condition, driving pulse intensity, THz absorption and geometric factor etc. are all inclusive [29]: $d_{eff}$ is non-linear coefficient for THz generation in LN crystal, $n(\omega_L)$ and $n(\Omega_{THz})$ are the phase index of the driving laser and THz in LN respectively, $\Delta k$ is the wave-vector mis-match in-between these two waves, being related to the TL driving pulse duration $\tau$ as well (according to Eq. (11-13)), $L_{THz}$ is THz propagation distance in the crystal, $\alpha_{THz}$ is the THz absorption coefficient, $F$ is the driving beam fluence (which divided by the pulse duration $\tau$ is equal to the peak intensity of the driving pulse), and $G(L_{THz}, \tau)$ is the geometric coupling function for the laser and THz beams, which is correlated to wave-front tilt geometry, $L_{THz}$ and $\tau$.

Utilizing the parameters in our unique experimental setup and wave-front tilt scheme, a series of simulated figures demonstrating "THz conversion efficiency"-$\eta$ vs. "driving pulse duration"-$\tau$ and "THz propagation length"-$L_{THz}$, for various absorption coefficients are plot in Fig.4. Due to the fact the refractive index of the media in THz band is not precisely characterized while the THz chromatic dispersion coefficient is expected to be small, it's reasonable to carry out the simulation using parameters at ~1THz (the refractive index of MgO:LN claimed to be 5.18~5.20) [17, 27, 30, 31]. The media absorption coefficient in THz band is not well calibrated either, but the general features are: i) it becomes smaller at low temperature (especially at the cryogenic environment); ii) the absorption becomes severely worse at higher THz frequency that is why THz crystals typically have cut-off frequencies (e.g. the cut-off frequency for LN is around ~1.5 THz); iii) excessive THz absorption is due to the laser induced free carriers, thus moderate laser fluence should be applied to reduce the effect [32, 33]. The distributions of THz generation efficiencies are highlighted for several typical THz absorption coefficients, where the extreme conversion efficiency for each case are very different, associated with diversified optimal driving pulse lengths and THz generation lengths. For THz absorption coefficient of LN is $\alpha_{THz}$~5cm$^{-1}$ (e.g. at room temperature), the optimal driving pulse duration is a bit below 100fs, the corresponding optimal THz generation length is 1~2mm, and the highest THz conversion efficiency is around ~0.15% (Fig. 4(a)), which is quite consistent with the previous reports for deriving THz wave by ~100fs pulse in LN at room temperature [26, 27]. If the absorption coefficient decreases to one order less, $\alpha_{THz}$~0.5cm$^{-1}$, the optimal pulse length should be around 400fs, with THz generation length beyond 30mm, more remarkably the extreme conversion efficiency is expected to be ~10% (Fig.4 (d)). During the past few years, cryogenic apparatuses were proposed and developed for THz generation, which could enhance the generation efficiency enormously because of the absorption is much smaller compared to that at room temperature [24, 26]. In that circumstance, an extra-large size pulse-front tilted driving beam with high pulse energy and long pulse duration (to maintain the beam fluence in few mJ/cm$^2$) should be applied to a cryogenic LN crystal with dimensions of few tens of mm, to achieve the desired THz conversion efficiency of a few percent. Currently one of the prevalent views for THz generation in LN through wave-front-tilt scheme is that the optimal driving pulse length should be in the range of 300-500fs unconditionally [34, 20, 26, 24]. However our investigation unveiled a different feature: the absorption coefficient of the media actually plays an essential role, and the optimal driving pulse length starts from ~400fs for small absorption coefficient ($\alpha_{THz}$~0.5cm$^{-1}$), decreases down to <100fs for large absorption coefficient ($\alpha_{THz}$~5cm$^{-1}$), while the optimal THz generation length and conversion efficiency decline 1~2 orders associatively. This could also well interpret previously unexpected experimental results at room temperature, where the THz conversion efficiency at the identical driving beam fluence decreases monotonously along with increasing of the pulse length, contradictory with the expectation value of the optimal pulse length in the range of 300-500fs [27, 20]. The optimal parameters for THz generation at various absorption coefficients are retrieved from Fig.4 and summarized in Table 1.

**Table 1.** Optimal driving pulse duration $\tau$, THz generation length $L_{THz}$ and extreme conversion efficiencies $\eta$ for various THz absorption coefficients $\alpha_{THz}$ in LN.

| $\alpha_{THz}$ [cm$^{-1}$] | Pulse Duration '$\tau$' [fs] | $L_{THz}$ [mm] | $\eta$ |
| --- | --- | --- | --- |
| 0.5 | ~380 | ~31 | ~10.0% |
| 1.0 | ~215 | ~14 | ~2.80% |
| 2.0 | ~140 | ~7 | ~0.90% |
| 5.0 | ~85 | ~2 | ~0.15% |

## IV. Experiment and Discussion

### A) Red-shift for IR residual spectra induced by cascaded DFG and THz pulse generation in non-linear regime

The generated THz is expected to escape normally from the side surface of 32.1mm x 15.0mm (X) of the LN crystal (refer to Fig.1). In the meantime, the residual driving laser beam via multiple internal reflections would exit from the same surface, but with a small deviation angle with respect to the THz beam. The residual beam diverges much quicker horizontally than vertically, due to the presence of tangentially chromatic and angular dispersion. A concave mirror is placed at an appropriate distance out of the exiting surface to collect the residual beam to feed into a spectrometer (Ocean Optics). The typical residual spectra of the driving beam fluencies in the range of 0.8-3mJ/cm$^2$ are plot in Fig.5. With increasing of the beam fluence, the spectra displaying significantly "red-shift" broadening features (usually claimed as the signature of "optical-to-THz" conversion with high efficiency) are successfully reproduced via our unique setup. However when the beam fluence increases to beyond 2.5mJ/cm$^2$, the trend of spectral 'red-shift' broadening slows down, indicating the saturation of the THz generation. THz wave is generated through the optical rectification in the crystal, with conversion efficiency much higher than "Manley-Rowe limit", where a driving photon converts into only a single THz photon and another optical photon of a small red-shift [24]. As illustrated in the inserted diagram of Fig.5, multi-stage cascaded DFG is expected to take place, thus an optical photon could convert into multiple THz photons through this cascaded frequency rolling down process.

The THz pulse energy is recorded and calibrated in the meantime by a pyroelectric detector located close to the crystal. With the driving beam fluence in the range of 0.06-3mJ/cm$^2$ (corresponding to the laser pulse energy of 0.015-1.2mJ), the generated THz pulse energy is from a few nano-Joules (nJ) up to micro-Joule (μJ) level (refer to Fig.6 (a)). The correlation of "Optical-to-THz conversion efficiency" vs. "Laser beam fluence" is plot in Fig.6 (b), and the efficiency is monotonously increasing and achieving ~0.09% for the highest beam fluence in the experiment, however when the fluence is beyond ~1.2mJ/cm$^2$, the THz generation is towards saturation gradually. The corresponding laser beam intensity is also given in the top-axis of Fig.6 (b), for the TL pulse duration of ~130fs. The experimental results of THz generation efficiency could be fit by a polynomial of $\eta(F) = c_0 + c_1*F + c_2*F^2 + \cdots$ (where F and $\eta$ are the laser beam fluence and THz generation efficiency respectively, and $c_j$ are the fitting coefficients), within two separated sections of the driving beam fluence, i) 0.06-1.2mJ/cm$^2$, ii) 1.2-2.8mJ/cm$^2$ respectively. The fitting parameters for these two sections are given in Table.2, where $c_1 \gg c_2$ indicates the THz generation is approximately in the linear conversion regime, and the non-negligibility of $c_2$ implicates that the THz generation is actually associated with a rather complicated process. Since $c_2$ is negative for the second order polynomial, the maximal THz generation efficiency is expected to achieve at the beam fluence of $-c_1/2c_2$. This simple scheme would lead to the highest conversion efficiency of 0.15% at the extrapolated fluence of ~5.32mJ/cm$^2$, for the lower beam fluence range (0.06-1.2mJ/cm$^2$); and to the conversion efficiency of 0.097% at ~3.90mJ/cm$^2$, for the higher beam fluence range (1.2-2.8mJ/cm$^2$). Both are clearly demonstrated by the extrapolated dash-curves in the plot (Fig.6 (b)). When the laser beam fluence is high, it would induce plasma distribution (or free carriers) in the crystal, which causes strong absorption of the THz wave; in the meantime, the plasma would lead to transient change of the refractive index of the media which potentially perturbs the phase match condition. So the attempt to enhance the THz generation efficiency via further increasing the driving beam fluence would be severely restricted, i.e. impossible. Then the saturated driving beam fluence of the experiment (corresponding to where the THz generation efficiency is maximal) is expected to be somewhere in-between the extrapolated values obtained from these two "low" or "high" fluence regions. In Fig.5, the overwhelmed 'red-shift' broadening feature is actually companied by 'blue-shift', mediated by the laser induced phase-modulation and broadband emission. And a direct comparison in-between the saturation trend for driving fluence of ~2.5mJ/cm$^2$ in the residual spectra (Fig. 5) and that of ~1.2mJ/cm$^2$ in THz generation efficiency curve (Fig.6), clearly shows the spectral broadening, not only corresponds to the feature of THz generation, but also includes that of the plasma induced by the laser, so the magnitude of 'red-shift' broadening in residual spectrum shouldn't be simply regarded as the THz generation efficiency. According to the analytic scheme used in Fig.6, the highest achievable THz generation efficiency in LN at room temperature by utilizing our apparatus and driving pulse length of ~130fs would be around ~0.1%, the order of magnitude is consistent with the feature predicted by Fig. 4(a) too.

**Table 2.** Fitting parameters for THz conversion efficiency.

| Coefficient | $c_0$ | $c_1$ | $c_2$ |
|---|---|---|---|
| Driving Beam Fluence [mJ/cm$^2$] | Unit of Coefficients | | |
| | [%] | [cm$^2$/mJ] | [cm$^4$/(mJ)$^2$] |
| 0.6-1.2 | 0.00153±0.00046 | 0.05744±0.00192 | -0.00540±0.00150 |
| 1.2-2.8 | 0.021361±0.00700 | 0.03903±0.00727 | -0.00501±0.00181 |

**B) THz time-domain waveforms derived by chirp pulses and the chirp influence on the THz generation efficiency**

The frequency of the THz waveform created in the experiment is measured by the traditional EO sampling method [11], where laser beam fluence of ~2.5mJ/cm$^2$ is used to derive THz wave via cascaded DFG in LN. The experimental EO signals representing typical THz fields in time domain along with its Fourier transformations in frequency domain are displayed in Fig.7 (a) and (b) respectively. Similar to the previous results of using LN, the generated THz spectrum spans from ~0.1THz up to ~1.5THz, with peak strength at 300-400GHz. Moreover, we stretched the driving laser pulse length up to 600~800fs, and applied the chirp pulses to excite the THz wave in LN. Fig.7 compares the normalized THz fields derived by TL pulse of 130fs and by chirp pulses of 250fs, -320fs in pulse length (where the positive or negative sign represents the positive or negative chirp). Apparently the THz waveform generated by chirp pulses (250fs or -320fs) is longer than that by TL pulse, thus corresponding to a narrower THz spectrum (it's reasonable to assume that the generated THz wave is close to full coherence). This indicates that at the identical intra-band distribution and beam fluence, the TL driving pulse with respect to chirp pulses, could generate a broader THz emission band due to its relatively higher conversion efficiency for generation of higher THz frequency.

In order to investigate the potential influence of the chirp of the driving pulse on the THz conversion efficiency, we generated different chirp pulses through slightly deteriorating the optimal compensation of the pulse compressor in the laser system. The pulse length is on-line characterized by a single shot auto-correlator (Coherent Inc., USA) via splitting few percent of the driving beam. Based on the TL pulse duration of ~130fs, the stretched chirp pulse parameters used in the experiment are summarized in Table.3. Under the linear chirp approximation, the general formula for the electric field of a chirp pulse could be expressed as,

$$E_N(t) = E_0 e^{-|a_N|t^2} \cos(\omega_c t + b_N t^2) \quad (15).$$

Where $\omega_c = 2\pi \frac{c}{\lambda_c}$ is the carrier frequency of the laser (at the central wavelength), coefficient $a_N$ and $b_N$ are related to the envelope duration and the linear chirp for the laser field respectively. In terms of TL pulse duration of FWHM $\tau_{FWHM}$ and pulse stretch multiple $N$ (the ratio of chirp pulse length divided by TL pulse length, sign included), we have [35],

$$a_N = sign(N) 2\ln 2 / \tau_{FWHM}^2 N^2 \quad (16.1),$$

$$b_N = a_N \sqrt{N^2 - 1} \quad (16.2),$$

$$1/\sqrt{|a_N|}^S = sign(N)/\sqrt{|a_N|} \quad (16.3).$$

Table 3. Parameters for various chirp pulses.

| Parameters | Pulse length stretch multiples, N | $a_N$ [fs$^{-2}$] | $b_N$ [fs$^{-2}$] |
|---|---|---|---|
| Negative Chirp | -3.65 | -6.153E-6 | -2.161E-5 |
|  | -1.47 | -3.821E-5 | -4.092E-5 |
| Chirp Free | ±1.00 | ±8.203E-5 | 0 |
| Positive Chirp | 1.10 | 6.866E-5 | 3.030E-5 |
|  | 1.33 | 4.668E-5 | 4.062E-5 |
|  | 1.67 | 2.926E-5 | 3.929E-5 |
|  | 1.74 | 2.696E-5 | 3.853E-5 |
|  | 2.07 | 1.915E-5 | 3.470E-5 |
|  | 3.14 | 8.322E-6 | 2.477E-5 |
|  | 6.22 | 2.120E-6 | 1.301E-5 |

Using Eq.(15-16) and implementing the parameters in Table 3, we plot the normalized field envelopes of the TL pulse of 130fs (green), negative chirp pulse of -475fs (red, N = -3.65) and positive chirp of 408fs (blue, N = 3.14) all together in Fig.8 (a). Especially within a time window of '70-130' fs respected to the peak field of the laser (semi-transparent bar area in the figure), the "Zoom-in" features of the field at the neighborhood of intra-pulse time scale of 70fs, 100fs and 130fs are highlighted for the three above pulses (in Fig.8 (b)). Obviously the TL pulse decays much faster than the chirp pulses, while their phase differences are discernable and increasing when the intra-pulse time is further away from the peak field.

According to the parameters in Table.3 and in Fig.8, the THz generation efficiency is systematically investigated for various chirped driving pulses while maintaining the pulse fluence as a constant at ~2.5mJ/cm². As demonstrated in Fig.9, a surprisingly asymmetric feature in the curve of "THz conversion efficiency" vs. "Chirp" is observed, where the signed inverse square root of the absolute value of $a_N$ (refer to Eq. (16.3)) is used to denote the chirp magnitude. And the highest conversion efficiency is achieved at a positive chirp value, not at the TL pulse. Especially when the driving pulse length is approaching to TL duration, the THz conversion becomes pretty low instead, but immediately increases to one order higher when positive chirps are applied; for negative chirps, the growth rate of the curve is moderated associated with apparently smaller magnitude of generation efficiency. As the pulse duration is further stretched, the THz generation efficiency gradually decreases and becomes more or less balanced for both sides of the chirp. The longest chirp pulse durations in the experiment are -475fs and 808fs for the negative and positive chirps respectively, due to the limit of the motor travel range of the pulse compressor.

### C) More discussions regarding to THz generation using chirp driving pulses

We developed a theoretical model to investigate the unexpected phenomenon of chirp asymmetry for the THz generation in wave-front tilt scheme. Concerning of a Gaussian driving pulse with linear chirp, the electric field in frequency domain could be expressed as

$$\tilde{E}(\omega) = E_0 \exp\left[-\frac{(\omega+\omega_c)^2}{4a_1} - ib_\omega(\omega+\omega_c)^2\right] \quad (17).$$

Where the parameter $a_1$ is equally defined as the series of $a_N$ in Eq. (16), for the stretching multiple N=1; $b_\omega$ represents the linear chirp parameter in frequency domain. Utilizing of the DFG simulation strategy in wave-front tilt scheme, we have,

$$\int_{-\infty}^{+\infty} d\omega' \tilde{E}_L(\vec{r},\omega') \tilde{E}_L^*(\vec{r},\omega'-\omega+\Omega_{THz}) e^{-i\vec{k}_{THz}\cdot\vec{r}}$$

$$= \int_{-\infty}^{+\infty} d\omega' |\tilde{E}_L(\vec{r},\omega')| e^{i\varphi(\omega')} |\tilde{E}_L^*(\vec{r},\omega'-\omega+\Omega_{THz})| e^{-i\varphi(\omega'-\omega+\Omega_{THz})} e^{-i\vec{k}_{THz}\cdot\vec{r}}$$

$$\approx \int_{-\infty}^{+\infty} d\omega' |\tilde{E}_L(\vec{r},\omega')|^2_{\omega=0} e^{i\varphi(\omega')-i\varphi(\omega'-\omega+\Omega_{THz})} e^{-i\vec{k}_{THz}\cdot\vec{r}} \approx \int_{-\infty}^{+\infty} d\omega' |\tilde{E}_L(\vec{r},\omega')|^2_{\omega=0} e^{-i\frac{\partial\varphi}{\partial\omega}\Omega_{THz}} e^{-i\vec{k}_{THz}\cdot\vec{r}}$$

(18).

The phase factor of Eq. (18) could be evaluated as the 'ω' terms in various orders,

$$-\frac{\partial \varphi(\omega)}{\partial \omega}\Omega_{THz} - \vec{k}_{THz} \cdot \vec{r} \approx -2b_\omega \omega \cdot \Omega_{THz} - \vec{k}_{THz} \cdot \vec{r} - \frac{z \cdot \Omega_{THz}}{k_L c^2} n(\omega - \omega_L)\left[n + \frac{\partial n}{\partial \omega}(\omega - \omega_L)\right]$$

$$= \frac{r \cdot \Omega_{THz}}{c}\left(n - \frac{\partial n}{\partial \omega}\omega_L\right) - \frac{z \cdot \Omega_{THz}}{c} n(\Omega_{THz}) + \left[-2b_\omega \Omega_{THz} - \frac{z \cdot \Omega_{THz}}{\omega_L c}\left(n - 2\omega_L \frac{\partial n}{\partial \omega}\right)\right]\omega + \frac{z \cdot \Omega_{THz}}{k_L c^2}\left(-n\frac{\partial n}{\partial \omega}\right)\omega^2 + \cdots$$

(19).

When the output value (absolute) of Eq. (19) approaches to minimum (i.e. zero), it is corresponding to the phase match condition for the highest THz conversion efficiency. Thus various orders of 'ω' terms could be analyzed separately.

i) Zero order of ω

$$\left[r \cdot \left(n(\omega) - \frac{\partial n(\omega)}{\partial \omega}\omega_L\right) - z \cdot n(\Omega_{THz})\right]\frac{\Omega_{THz}}{c} = 0 \qquad (20.1).$$

It represents the primary phase-match condition for the THz generation, where r and z represent the travel distances of the laser and THz beams respectively. At the condition of $r = z/\cos\theta$, Eq. (20.1) is equivalent to the wave-front tilt scheme described by Eq. (2-4).

ii) First order of ω

$$b_\omega = -\frac{z}{2\omega_L c}\left(n(\omega) - 2\omega_L \frac{\partial n(\omega)}{\partial \omega}\right) \qquad (20.2)$$

This indicates applying a certain amount of linear chirp for the pump laser pulse would promote THz generation, and the desired linear chirp value is expected to be proportional to the optimal THz generation length – 'z' in the LN crystal.

iii) Second order of ω and above

$$\frac{z\Omega_{THz}}{k_L c^2}\left(-n(\omega)\frac{\partial n(\omega)}{\partial \omega}\right)\omega^2 + 0\left[\omega^3\right] \qquad (20.3)$$

It is related to the other non-linear complex processes in the THz generation, e.g. laser induced transient change of the refractive index, scattering and diffraction effect of the media and etc., but its contribution to the overall phase should be relatively smaller compared with the first two terms.

Then applying Fourier transform to Eq. (17), the linear chirp parameter of Gaussian pulse in time domain could be determined,

$$E_N(t) = \frac{1}{\sqrt{2\pi}} \int_{-\infty}^{+\infty} E_0 \exp\left[-\frac{\omega^2}{4a_1} - ib_\omega(\omega+\omega_c)^2\right] \cdot e^{i\omega t} d\omega$$

$$= \frac{1}{\sqrt{2\pi}} E_0 \cdot e^{-i\omega_c t} \int_{-\infty}^{+\infty} \exp\left[-\frac{\omega^2}{4a_1} - ib_\omega \omega^2\right] \cdot e^{i\omega t} d\omega = E_0 \cdot e^{-i\omega_c t} \sqrt{\frac{2a_1}{1+4ia_1 b_\omega}} e^{-\frac{a_1}{N^2} t^2} \cdot e^{+4i\frac{a_1^2 b_\omega}{N^2} t^2}$$

(21).

Where $N = \sqrt{1+16a_1^2 b_\omega^2}$ is identical to the stretching multiple in Eq. (17). Comparing the phase term in Eq. (21) (for the real part) with that in Eq. (15), the linear chirp in time domain $b_N$ could be correlated to the value in frequency domain $b_\omega$, whose optimal value is given in Eq. (20.2),

$$b_N = -4\frac{a_1^2 b_\omega}{N^2} \quad (22).$$

Combining Eq. (22), Eq. (20.2) and Eq. (16), the optimal stretching multiple for the chirp pulse, $N^{(OPT)}$ (where the superscript of 'OPT' denotes as the optimal value) could be determined through,

$$\sqrt{(N^{(OPT)})^2 - 1} = -\frac{2\ln 2}{\tau_{FWHM}^2} b_\omega^{(OPT)} = \frac{z \cdot \ln 2}{\tau_{FWHM}^2 \omega_L c}\left(n_L - 2\omega_L \frac{\partial n_L}{\partial \omega}\right)$$

$$= \frac{z \cdot \lambda_L \cdot \ln 2}{\tau_{FWHM}^2 c^2}\left(n_L + 2\lambda_L \frac{\partial n_L}{\partial \lambda}\right)$$

(23).

Obviously $N^{(OPT)}$ is dependent with the THz generation length –'z' as well. Implementation the parameters in the experiment, we obtain the optimal chirp pulse stretching multiples are $N^{(OPT)}$ = 1.43 or 2.31 for the THz generation length of 1 or 2 mm respectively, associated with the calculated value of $1/\sqrt{|a_N|}^S$ in the range of 158-255fs. And the experimental value of N~2.07 for the optimal THz conversion efficiency (i.e. $1/\sqrt{|a_N|}^{S(OPT)}$ ~228 fs) is well located inside this regime (refer to Fig.9).

Since the laser pulse fluence of ~2.5mJ/cm$^2$ used in the experiment is beyond the THz linear conversion region, therefore the conversion efficiency is relatively low at the TL pulse duration. And application of the positive chirp for the driving pulse would promote the THz conversion efficiency dramatically. Since the chirp parameters e.g. $a_N$, $b_N$ are signed values, the positive and negative values display completely different feature for satisfying the optimal THz generation condition (i.e. Eq. (23) is only valid for the positive 'N', but not for the negative one), so the chirp asymmetry behavior is sensible. And further theoretical exploration with thorough details would appear in our future publications.

**V. Conclusion**

Here we developed a novel THz generation apparatus by utilizing a unique 'dual-face-cut' LN crystal in Brewster's incidence coupling geometry, and experimentally demonstrated the optical-to-THz conversion efficiency of ~0.1% could be achieved at room temperature by using conventional Ti:Sapphire laser (~800nm) with pulse length of ~130fs. This feature is similar to the typical cascaded THz generation using wave-front tilt scheme in 'z-cut' LN crystal. Since the material chromatic dispersion of LN in the neighborhood of the driving wavelength is small, our THz generation scheme could potentially be extrapolated to broader driving wavelength range as well, with only minor tuning required.

We systematically investigate the various determinants which restrict the THz conversion efficiency, and recognized that the chromatic dispersion and angular distribution are unavoidable for the wave-front tilt scheme, corresponding to the most primary constrains for improving the THz generation length and conversion efficiency. And we found out that the THz absorption coefficient $\alpha_{THz}$ of the media plays an essential role, when it decreases from ~5cm$^{-1}$ (at room temperature) to below ~1cm$^{-1}$ (probably at the cryogenic temperature), the THz conversion efficiency would be increased to more than one order higher, from ~0.1% into few percent level, and the extreme conversion efficiency is predicted at ~10%. In order to achieve this, the optimal driving pulse duration should be increased from ~100fs up to ~400fs and the corresponding optimal THz generation length should be enhanced from 1~2mm up to few tens of mm.

Furthermore, the influence of the chirp of the driving pulse on the THz generation efficiency is studied, for the identical driving pulse fluence of ~2.5mJ/cm$^2$ at the mild saturated region. We discovered that the THz conversion efficiency is not optimal at the TL pulse length, instead the positive chirp would promote much higher conversion efficiency (1 or 2 orders higher) compared to the negative chirp, exhibiting a surprising asymmetry feature in the conversion efficiency curve. We developed a simulation script to interpret this intriguing feature, and reproduced the experimental results successfully. According to our study, applying appropriate positive chirp for the driving pulse could potentially be a general strategy to enhance THz generation efficiency in wave-front-tilt scheme.

From Eq. (23), the desired magnitude of the chirp should be proportional to $\propto z / \tau_{FWHM}^2$. The optimal THz generation length is few tens of mm (e.g. ~31mm), and the optimal pulse duration is ~400fs for the small absorption coefficient case (e.g. $\alpha_{THz}$~0.5cm$^{-1}$), comparing with ~2mm and ~100fs for the large absorption case (e.g. $\alpha_{THz}$~5.0cm$^{-1}$). Therefore the optimal stretching multiple N$^{(OPT)}$ for both cases should be more or less similar. However the long TL pulse length of ~400fs is associated with a much narrower intra-pulse bandwidth compared to the short pulse of ~100fs, so it would require a pulse stretcher or compressor to create relatively larger intra-pulse dispersion to provide the desired chirp value. This might be quite useful for the experimental practice by using a TL pulse length in 300-400fs to derive THz wave in cryogenic setup, and it is worthwhile to generate chirps for the driving pulses and investigate if THz generation efficiency could be enhanced as expected. Due to the lack of appropriate laser systems providing 300-400fs TL pulses in our laboratory, we are not able to investigate this directly, but more delicate apparatus are currently under construction for further investigation.

**Acknowledgement**

The work is partially supported by the National Science Foundation of China (grants # 11475249), and Youth 1000-Talent Program in China (grants # Y326021061). The authors thank for the staff and facility support from the Dept. of Free Electron Laser Science & Technology, and Water Science Research Center, Shanghai Institute of Applied Physics, and support from Laboratory for High Intensity Optics, Shanghai Institute of Optics and Fine Mechanics. The authors also appreciate the previous support and valuable discussion with Prof. Kaertner and group at DESY, Hamburg.

**Figures and Figure Captions**

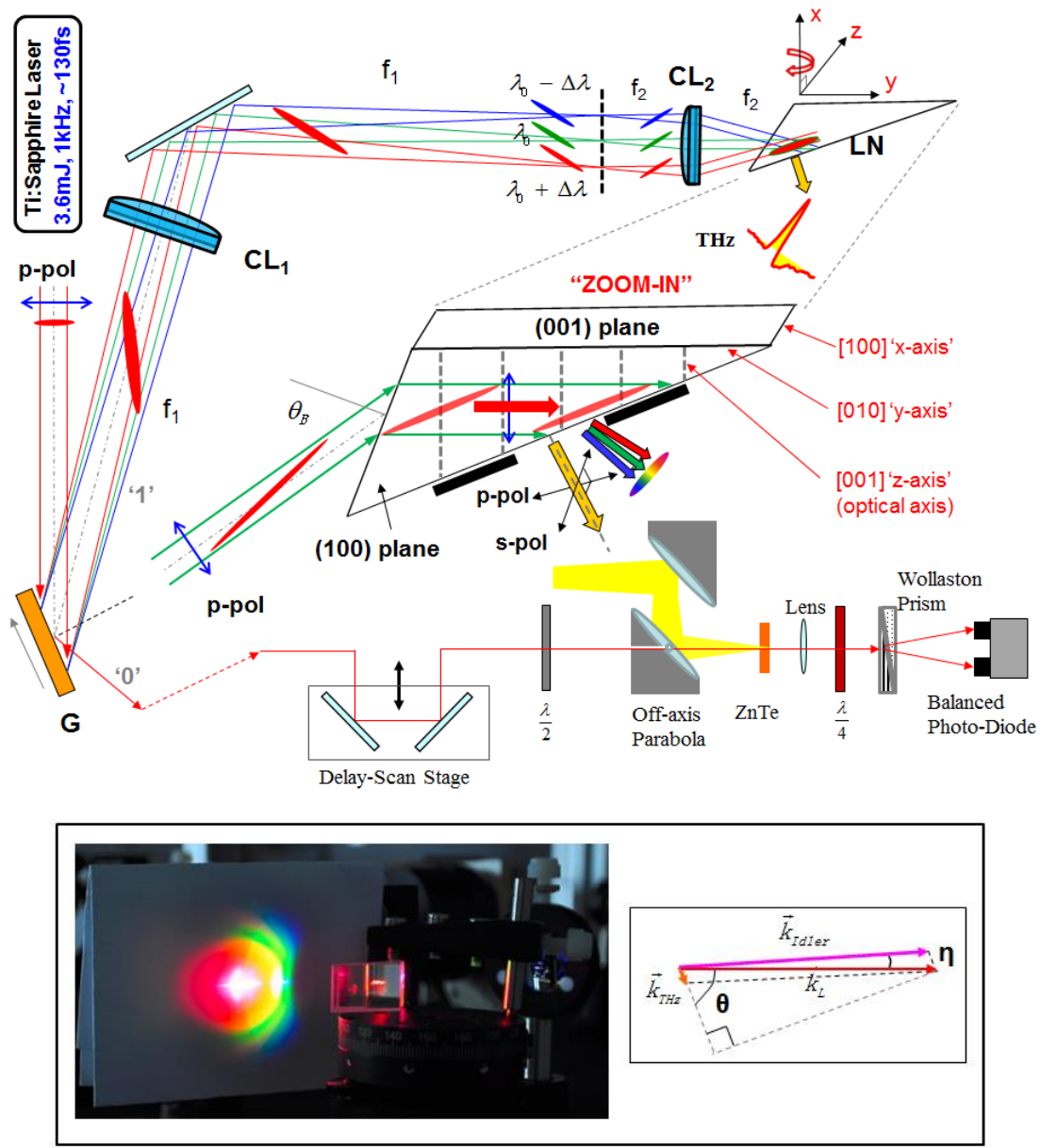

**Figure 1** The schematic layout of the THz generation setup ("bird-view"), implementing the novel design of wave-front tilt scheme through Brewster incidence coupling. The center of the figure shows the "Zoom-in" 3D-feature of the wave-front tilted driving beam propagation through the LN crystal. The 1st order - diffraction beam ('1') of the grating (G) possessing wave-front tilt is used to derive THz wave in the LN crystal, and the 0th order - reflection beam ('0') is used as probe to characterize the THz field in time domain through EO sampling technique. The diffraction beam is delivered through a pair of cylindrical lenses (CL$_1$/CL$_2$) to obtain a desired demagnification of f$_1$/f$_2$ in its tangential coordinate before fed into LN. The inserted photograph and vector diagram underneath illustrate the plasma enhanced THz generation in the crystal and the phase matched non-collinear DFG process for achieving high "optical-to-THz" conversion efficiency (see text).

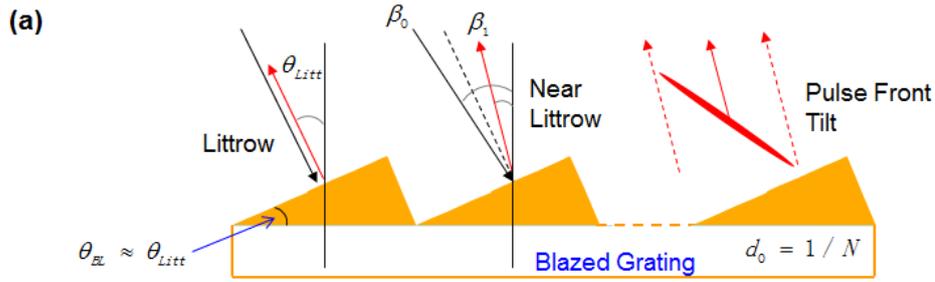

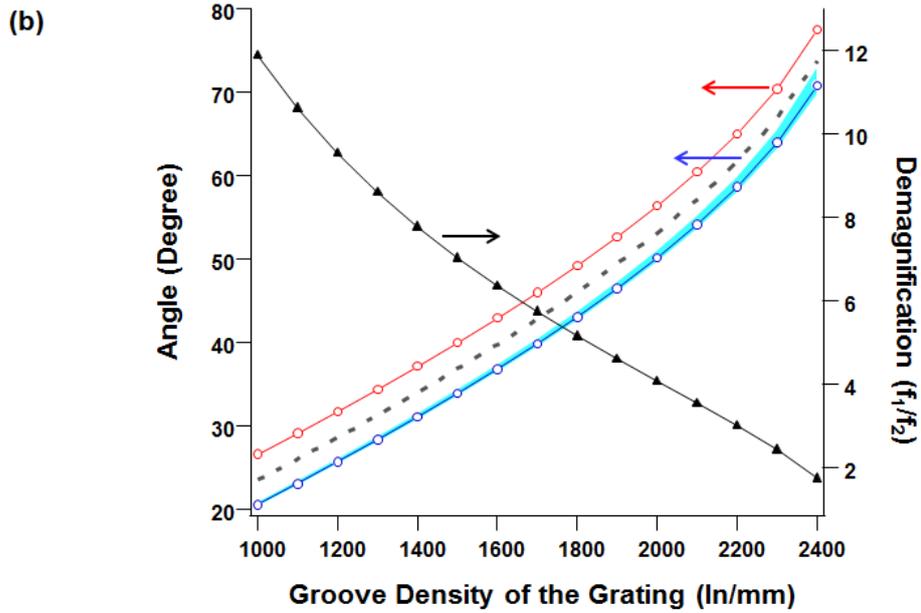

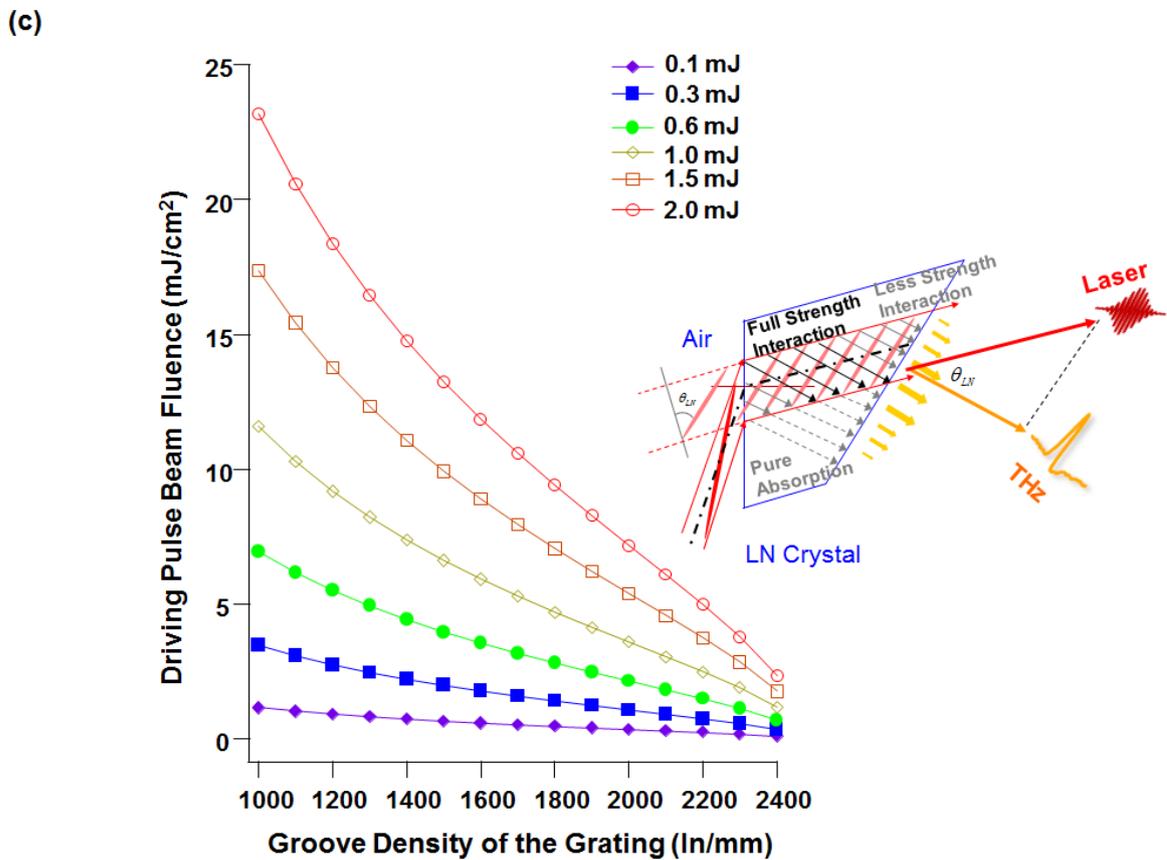

**Figure 2 (a)** illustrates the wave-front tilt generation for the laser beam through a blazed grating, with blazed angle equal to 'Littrow' angle for the central wavelength. The incidence beam approaches the Littrow condition for high diffraction efficiency, and a few degrees' deviation from the Littrow angle could avoid spatial overlapping in-between the incidence/diffraction beams. **(b)** The proposed "Incidence" (red-circle-line) & "Diffraction" (blue-circle-line) angles, distributed near to the Littrow angles (grey-dash-line) for a series of blazed gratings with groove densities in the range of 1000-2400ln/mm (left axis). The shadow domain (light-blue) displays the angular distribution of the "Diffraction" beam due to the intra-band chromatic dispersion of the driving pulse. The optimal demagnification coupling of the lens-pair, $f_1/f_2$ (filled black-triangle, right axis) in the THz generation setup to achieve the phase match in the LN crystal for the series of gratings working close to Littrow condition (see text). **(c)** The expectation values of the driving pulse fluence for laser pulse energy in the range of 0.1-2.0mJ, using the same series of gratings in (b). The inserted schematic diagram demonstrates the wave-front tilted driving pulse penetration and interaction with the LN crystal to generate THz wave, whose transverse intensity distribution is due to the variety of the THz interaction path lengths with the laser wave-packet.

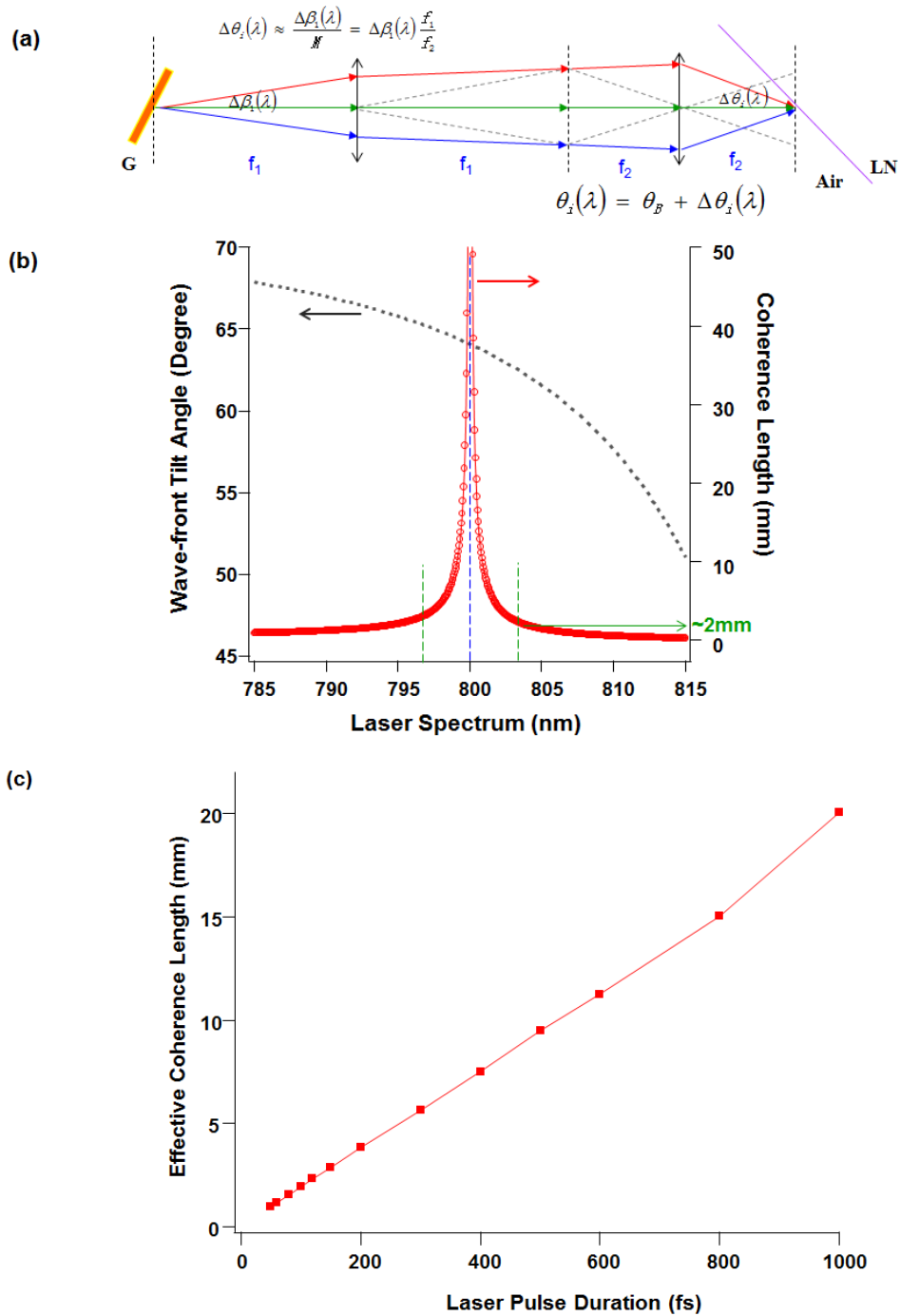

**Figure 3 (a)** The chromatic effect of the driving beam, where various intra-band wavelengths are associated with different optical paths, in-between the grating (G) and LN crystal, via a pair of tangential de-collimation lens-pair. **(b)** For the grating with groove density of 2200ln/mm and driving laser intra-bandwidth of 7.6nm at ~800nm, the distribution of the wave-front-tilt angles for specific wavelengths in LN, respected to the optimal phase-match angle of 64°at the central wavelength (dot-line, left-axis); and the coherence length for THz generation (circle-line, right-axis), where the two short vertical dash-lines indicate the intra-band angular dispersion in the experiment would lead to the declining of the THz coherence generation length down to ~2mm. **(c)** The plot of "effective coherence length" for THz generation vs. the transform-limit "pulse length" of the driving laser.

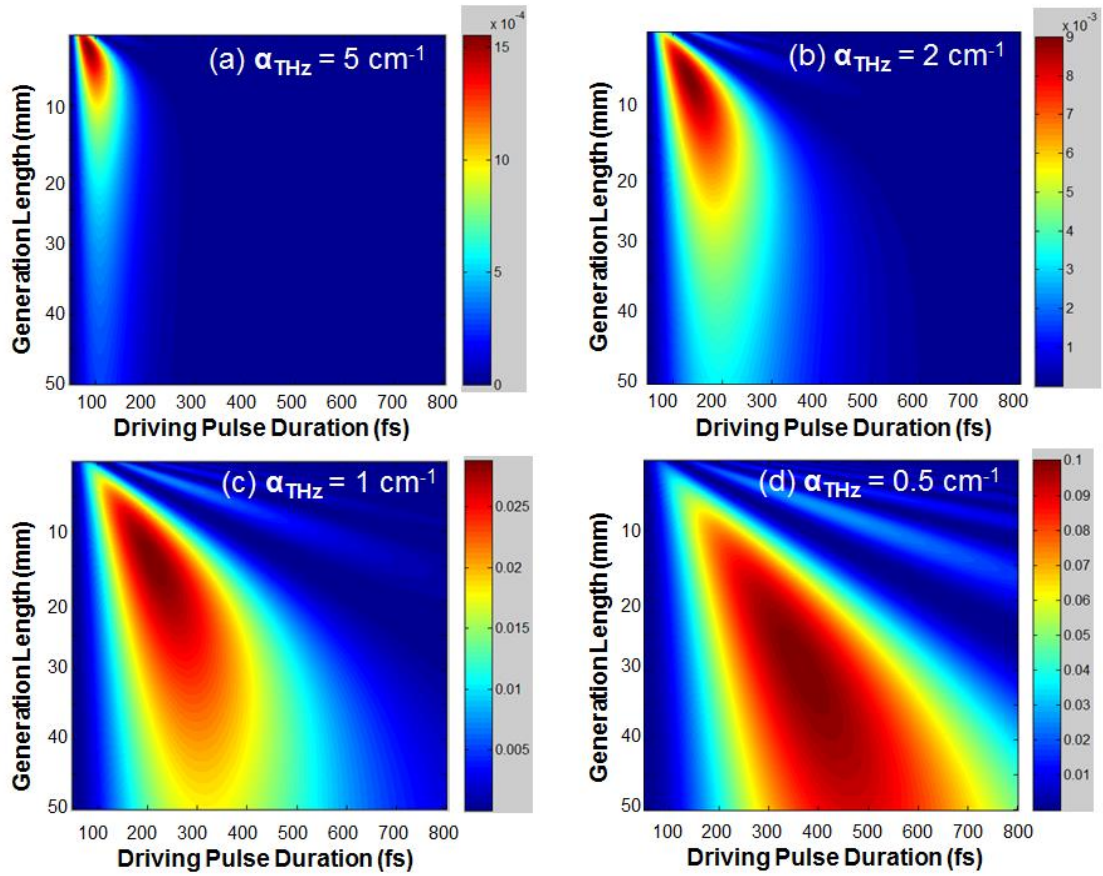

**Figure 4** The simulated THz generation efficiency for various TL "driving pulse duration" and THz "generation length", at different THz absorption coefficients "$\alpha_{THz}$" of the crystal, **(a)** α$_{THz}$ = 5.0 cm$^{-1}$, **(b)** α$_{THz}$ = 2.0 cm$^{-1}$, **(c)** α$_{THz}$ = 1.0 cm$^{-1}$, **(d)** α$_{THz}$ = 0.5 cm$^{-1}$. In each diagram, the absolute values of the THz conversion efficiency distribution are denoted by the color-bar, and the optimal parameters (i.e. "Driving pulse duration (TL)" & "THz generation length") could be identified and retrieved from the figure for each case (see Table 1).

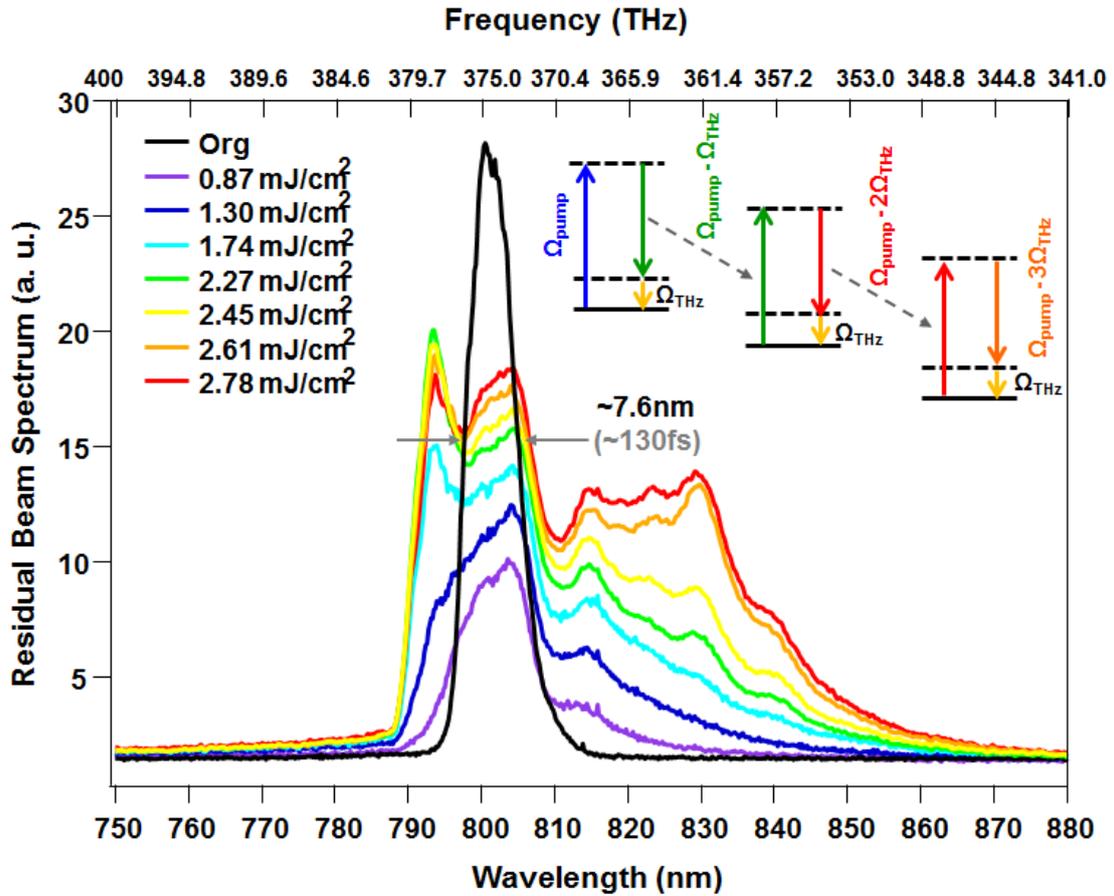

**Figure 5** The "red-shift" broadening of the residual spectrum of the driving laser for the THz generation in LN, at various beam fluencies. The black one is the original laser spectrum as the control signal. The inserted diagram shows the THz generation in LN is via multi-stage cascaded DFG process, causing the broadening feature in the residual spectra.

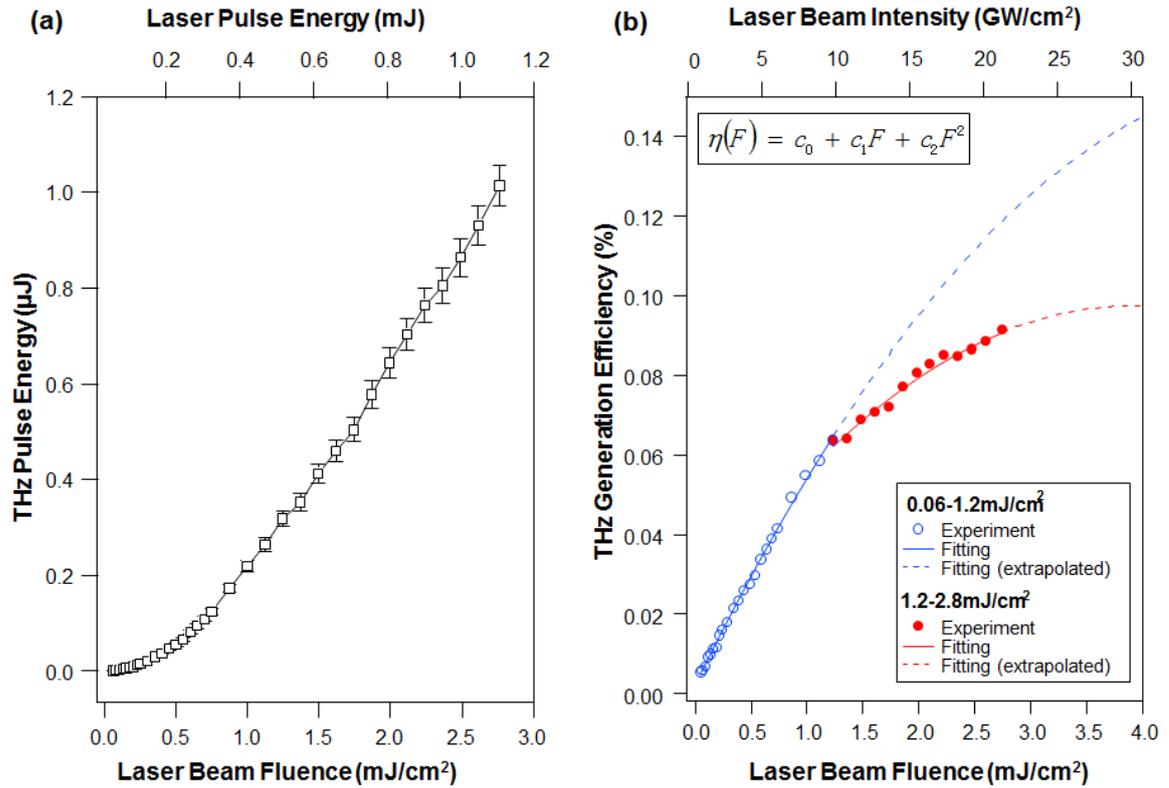

**Figure 6 (a)** The THz pulse energy generated in the experiment by applying different laser beam fluencies in the range of 0.06-2.8mJ/cm$^2$ (associated with the laser pulse energy of 0.015-1.2mJ). The error bar in each data point indicates that the uncertainty for charactering the THz pulse energy is less than 5%. **(b)** The corresponding THz generation efficiency, calculated by using experimental data in (a). The domains of the low (open-circle) and high (filled-circle) beam fluencies could be fit by polynomials of second order (blue and red solid-lines), achieving two sets of parameters (in Table 2). The dash curves in specified colors are the extrapolated features of the polynomials for these two domains (see text).

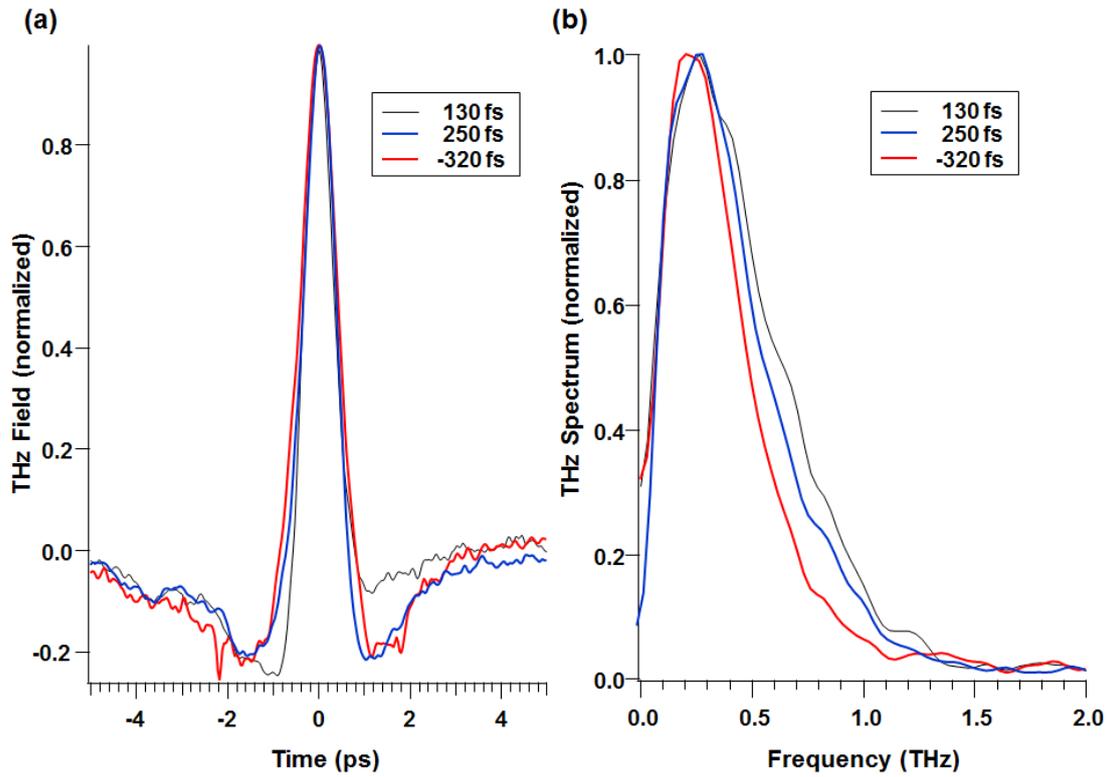

**Figure 7 (a)** The measured THz waveform in time-domain by electric optic sampling, the thin black curve is generated by the TL driving pulse of 130fs, and the blue and red lines by positive chirp pulse of 250fs and negative pulse of -320fs respectively. **(b)** The corresponding THz spectra in frequency-domain, via Fourier transformation of (a).

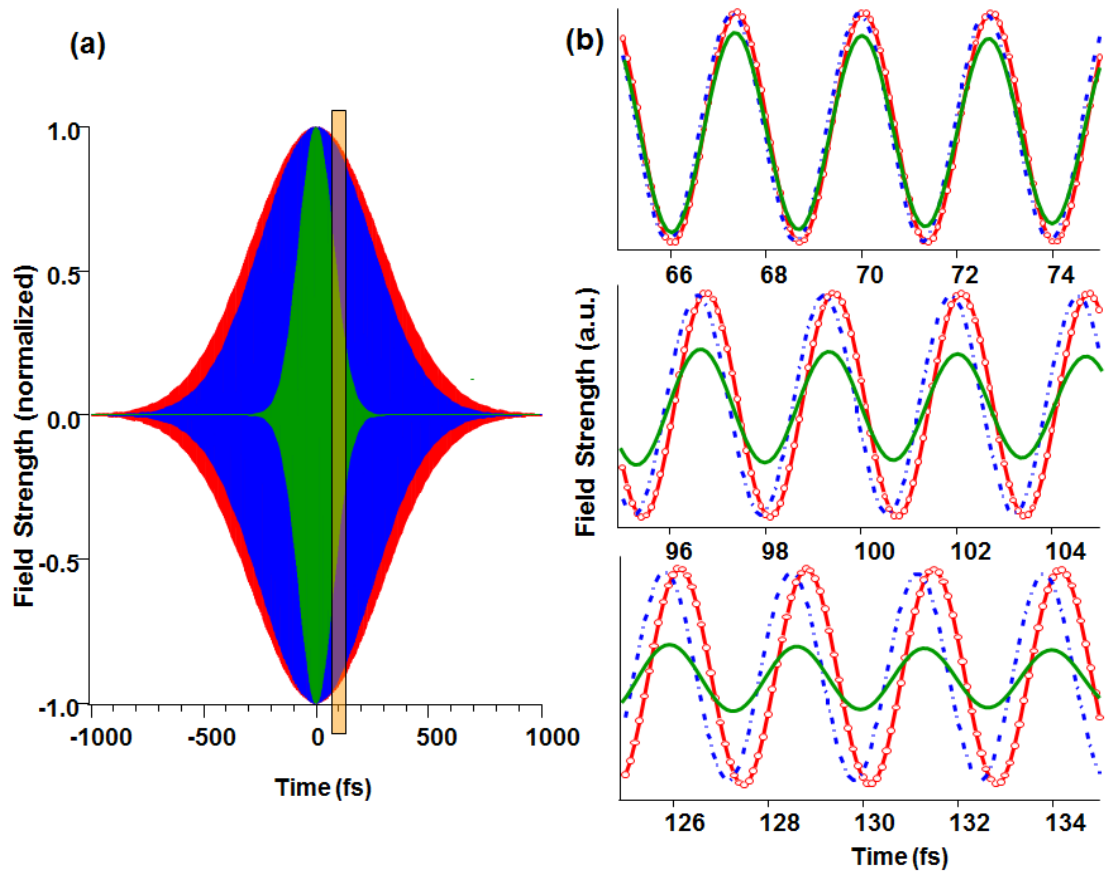

**Figure 8 (a)** The simulated driving pulse waveforms (normalized) for TL pulse duration of 130fs (green), negative chirp pulse duration of -475fs (red), and positive chirp pulse duration of 408fs (blue) used in the experiment. The vertically semi-transparent bar-area across the three wave-packets is representing a time-window of (65,135) fs with respect to the field peak. **(b)** "Zoom-in" views for three typical "10-fs" sub-windows within the "65-135 fs" window (the central time scales of the three sub-windows are 70, 100, 130fs with respect to the field peak), and comparison of the electric fields of the three normalized pulses, showing the phases of negative (red circle-line) and positive (blue dot-line) chirp pulses are shifting oppositely respected to the chirp-free pulse (green solid-line).

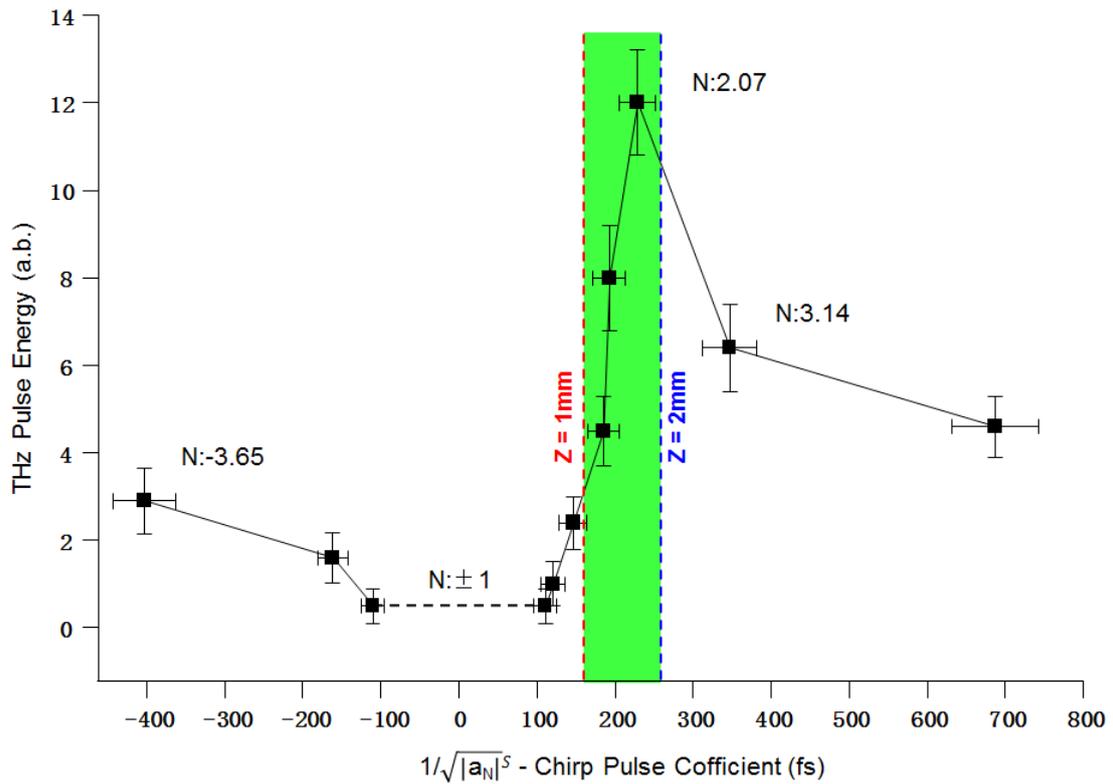

**Figure 9** The THz pulse energy generated by using identical driving beam fluence of ~2.5mJ/cm², while the chirp of the pulse is changed from negative, to chirp-free (corresponding to TL pulse), and then to positive. A few typical stretching multiples –'N' from the experiment are highlighted near to their corresponding data points. The two vertical dash lines and enclosed bar-area represent the spanning range of the calculated chirp pulse coefficients $1/\sqrt{|a_N|}^{S(OPT)}$ for the optimal THz conversion efficiency within the generation length range of 1-2mm, according to Eq. (23). Obviously the optimal value of stretching multiple $N^{(OPT)}$ ~2.07, associated with the highest THz generation efficiency is well located within the region (see text).